\documentclass[11pt]{article}
\usepackage{moriond,epsfig}

\def\be{\begin{equation}}
\def\ee{\end{equation}}
\def\bea{\begin{eqnarray}}
\def\eea{\end{eqnarray}}

\begin{document}
\vspace*{4cm}
\title{JET RECONSTRUCTION AT RHIC}

\author{ Sevil Salur for the STAR Collaboration}

\address{Department of Physics,
University of California, 
One Shields Avenue, Davis, CA 95616 USA}

\maketitle\abstracts{Full jet reconstruction in heavy-ion collisions is expected to provide
more sensitive measurements of jet quenching in hot QCD matter at RHIC.
In this paper we review recent studies of jets utilizing modern
jet reconstruction algorithms and their corresponding background
subtraction techniques.}

\section{Introduction}

Jets can be used to probe the properties of the high energy density matter  created in the collisions at the Relativistic Heavy Ion Collider (RHIC). Strong suppression of inclusive hadron distributions and di-hadron correlations at high $p_{T}$ have provided evidence for partonic energy loss in an indirect way \cite{starsub,phenixsub}. These measurements however suffer from well-known geometric biases due to the competition of fragmentation and energy loss.  It is possible to avoid the geometric biases if the jets are reconstructed independent of their fragmentation details  whether they are quenched or unquenched.  In this paper, we discuss the current status of the jet reconstruction in heavy ion collisions and the implication of the results.

\section{Jet Reconstruction Techniques}

During the last several decades, many algorithms were developed to combine measured particles into jets in leptonic and hadronic colliders.    For a detailed overview of jet algorithms in high energy collisions, see \cite{jetsref,seymor,jets,blazey,kt,ktref} and references therein.  Measuring jets above the complex heavy ion background however is a challenging task. For a long time, it has thought to be not possible due to the large underlying high multiplicity heavy ion event background.  The expected increase in Large Hadron Collider  (LHC) luminosities (20 to 200 collisions in a detector) leading to p+p pile up events requires that the traditional jet algorithms  are to be improved with underlying event subtraction techniques. These improved techniques can be also used to reconstruct and separate  jets from the underlying heavy ion background \cite{catchment}.

 The minimum requirement for an unbiased jet reconstruction in heavy ion collisions is that the signal and the background must be separable.  With the assumption that it can be,  the background correction can be estimated by following three steps.  The first step is measuring the jet area for the infrared safe algorithms. An active area of each jet is estimated by filling an event with  many very soft particles and then counting how many are clustered into a given jet. The second step is measuring the diffuse noise  (mean $p_{T}$ per unit area in the remainder of the event) and noise fluctuations.  These fluctuations in the background can distort the jet spectrum towards larger $p_{T}$ which can be corrected  through an unfolding procedure (i.e., deconvolution). So the final step is the deconvolution of signal from the background using parameters that are extracted from measurable quantities. 


\section{Results}

\begin{figure}[h]
\begin{minipage}{16pc}
\includegraphics[width=16.pc]{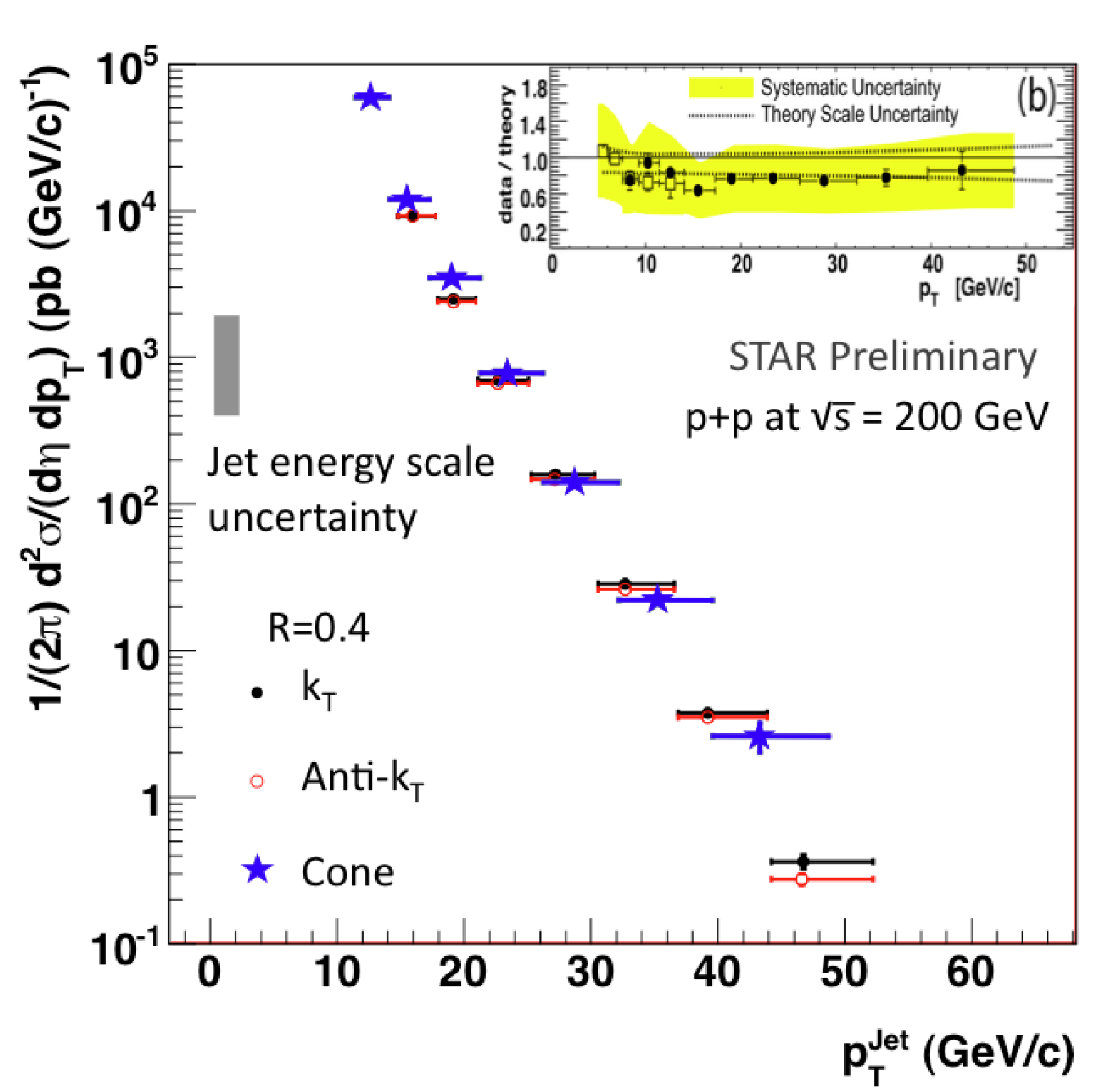}
\caption{\label{fig:spectra} Inclusive jet cross-section vs transverse jet energy for the p+p collisions obtained by the sequential recombination ($\rm k_{T}$ and anti-$\rm k_{T}$) algorithm  (shown as circles) and the previously published cone jets (shown as stars). Gray band is the jet energy scale uncertainty. Inset shows the comparison of the STAR cone jets with the NLO pQCD cross-section calculations.}
\end{minipage}\hspace{2pc}%
\begin{minipage}{17pc}
\includegraphics[width=16pc]{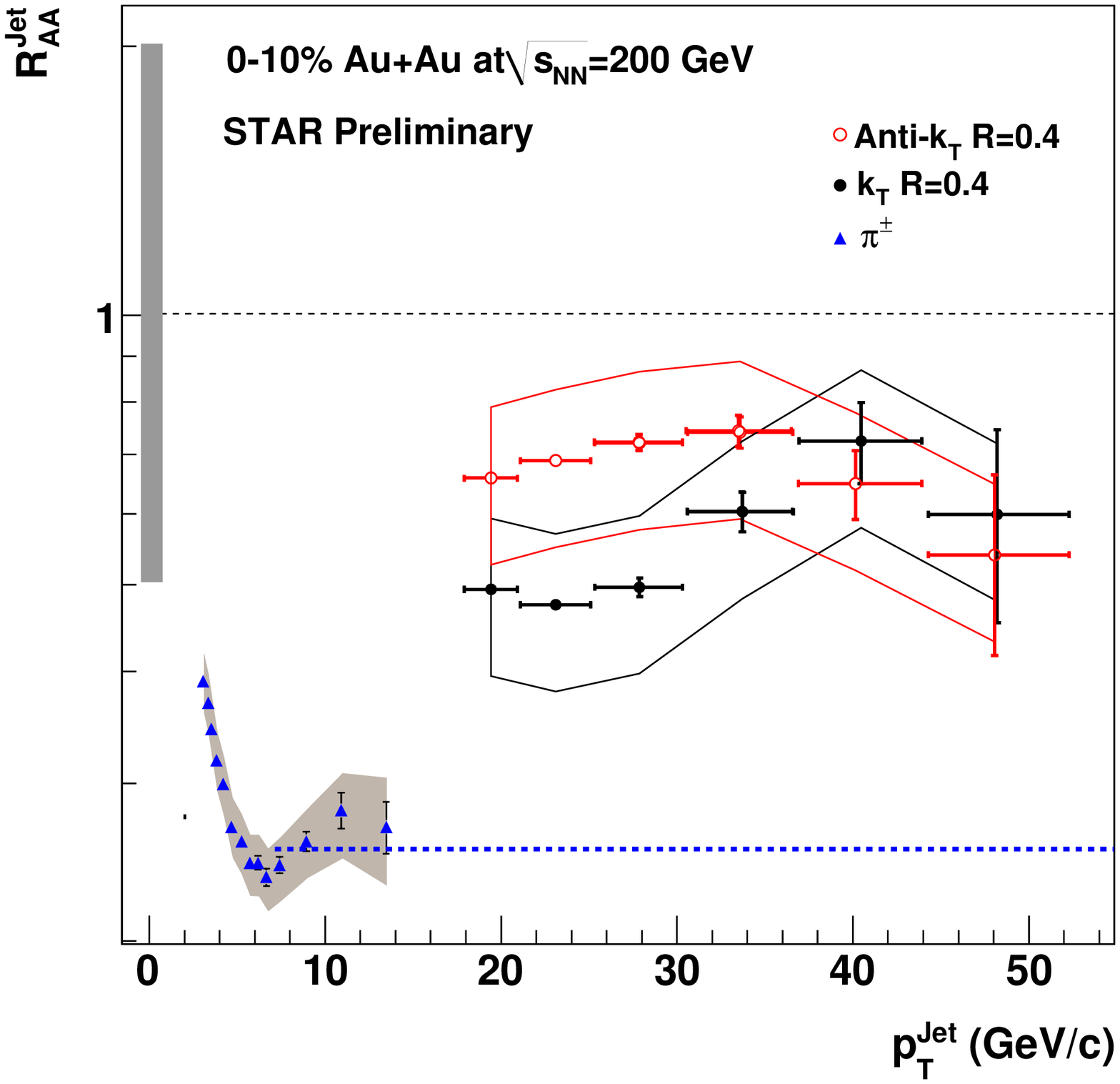}
\caption{\label{fig:raastar} Momentum dependence of the nuclear modification factors of jet spectra reconstructed with $ \rm k_{T}$ and anti-$\rm k_{T}$ algorithms with R=0.4 (0-10\% most central Au+Au divided by $\rm N_{Binary}$ scaled p+p collisions) compared to  $\pi^{\pm}$ $\rm R_{AA}$. The systematic uncertainty of the $\pi$ measurement is shown as the gray band and the gray bar centered at 1 is the jet energy scale uncertainty. The dashed lines are to guide the eye for $\rm R_{AA}=1$ and single particle $\rm R_{AA}$.   }

\end{minipage} 
\end{figure}

The transverse momentum dependence of the  inclusive differential cross sections for $p+p\rightarrow jet +X$ at $\sqrt{s}=$200 GeV are shown in Figure ~\ref{fig:spectra}.  The sequential recombination algorithm jets shown as circles using FastJet suite of algorithms are compared to jets  reconstructed with a cone algorithm as shown as blue stars  \cite{starpp,ploskon,fastjet}.  Both resolution parameters for  $\rm k_{T}$ and anti-$\rm k_{T}$ algorithms  and the cone radius are selected to be 0.4. 
These jet cross-sections agree well with each other within their statistical and systematic  uncertainties. The comparison of cone jets to  NLO pQCD cross-section using the CTEQ6M parton distributions is presented in the inset of Figure~\ref{fig:spectra} \cite{starpp,nlo}.   A satisfactory agreement for cross-sections over 7 orders of magnitude shows that jets in p+p collisions at RHIC energies are also theoretically well understood like the jets that are produced at the Tevatron energies \cite{tevatron}.

 The nuclear modification factor ($\rm R_{AA}$) for the reconstructed jet spectra with a resolution parameter of 0.4 from $\rm k_{T}$  and anti-$\rm k_{T}$ can be calculated after jets are reconstructed and corrected in Au+Au collisions. The preliminary version of the jet spectra in Au+Au collisions can be found in other publications \cite{ploskon,salurHP}.  Figure~\ref{fig:raastar} shows the $\rm R_{AA}$ of jets in Au+Au collisions.  The envelopes represent the one sigma uncertainty of the deconvolution of the heavy ion background. The total systematic uncertainty due the jet energy scale is around 50\%, shown as the gray bar.  The jet $\rm R_{AA}$ is compared to the one from the charged $\pi^{\pm}$ mesons \cite{yichunQM}. 
 
 In the case of full jet reconstruction, $\rm N_{Binary}$ scaling as calculated by a Glauber model \cite{glauber}  ($\rm R_{AA}=1$) is expected if the reconstruction is unbiased, i.e. if the jet energy is recovered fully independent of the fragmentation details, even in the presence of strong jet quenching. This scaling is analogous to the cross section scaling of high $p_{T}$ direct photon production in heavy ion collisions observed by the PHENIX experiment \cite{phenix}.    While the experimental uncertainties are large, a trend towards a much less suppression than that of single particle suppression is observed with the implication that a large fraction of jets are reconstructed. However a hint of a suppression of jet $\rm R_{AA}$ above 30 GeV can be observed. 
 
\begin{figure}[h]

\begin{minipage}{15.5pc}

\includegraphics[width=15.5pc] {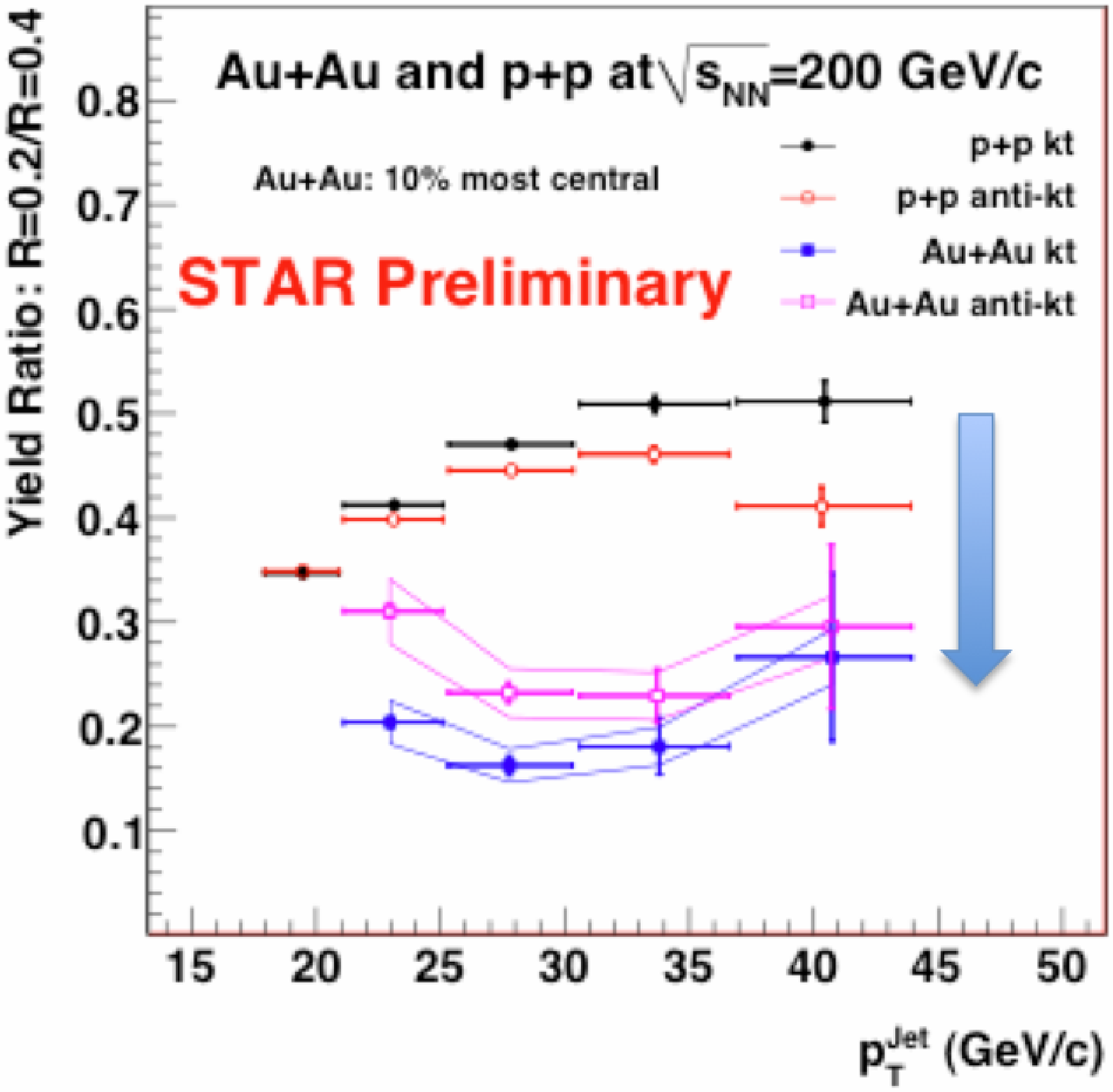}

\caption{ \label{fig:spectraratio} Momentum dependence of the ratio of inclusive jet cross-sections (R(0.2)/R(0.4)) reconstructed by $\rm k_{T}$ and anti-$\rm k_{T}$ recombination algorithms for p+p and Au+Au collisions.  } 
\end{minipage}\hspace{2pc}%
\begin{minipage}{17.5pc}
\includegraphics[width=17.5pc]{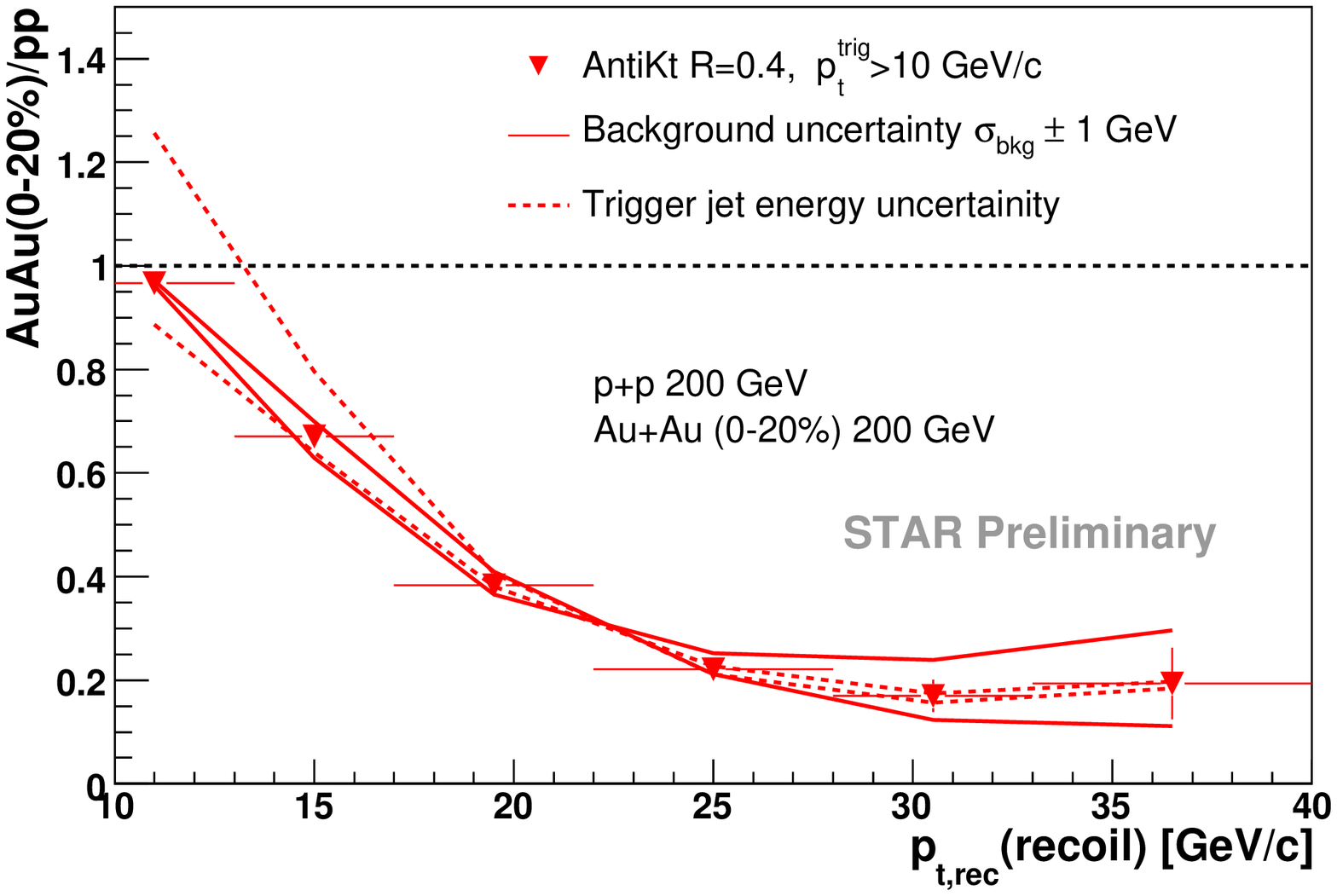}
\caption{\label{fig:dijet} Momentum dependence of the ratio of the spectra of the recoil jets  in 0-20\% central Au+Au to p+p collisions utillizing the HT trigger events. The systematic uncertainties in the estimation of the  background fluctuations and the triggered jet energy are shown as the solid and the dashed lines.}

\end{minipage}
\end{figure}

The ratio of jet spectra reconstructed with R=0.2 and 0.4 for p+p and Au+Au systems for $\rm k_{T}$ and anti-$\rm k_{T}$ is presented in Figure~\ref{fig:spectraratio}. A suppression in the Au+Au ratio with respect to p+p is observed. 
For a smaller resolution parameter due to possible additional jet broadening effects in Au+Au collisions, a larger fraction of the jet energy is not recovered unlike the jets reconstructed in p+p events.    The jet broadening effects can be investigated by selecting a biased sample of recoil jets in di-jet coincidence measurements.  The ratio of the spectra from the recoil jets in 0-20\% central Au+Au to p+p collisions \cite{brunaQM} is presented in Figure~\ref{fig:dijet}.  The recoil jets are selected when the triggered jets have $p_{T}$ greater than 10 GeV. Before taking the ratio, the recoil jet spectra in p+p and Au+Au collisions are normalized to the number of triggered jets.    When a population of recoil jets biased towards the ones that are interacting with the medium are selected, the effects of jet broadening can be observed to be much more comparable to the measurement of $\pi$ meson $\rm R_{AA}$.  This is in contrast to the inclusive jet measurements yielding a much smaller  nuclear modification suppression as seen in Figure~\ref{fig:raastar}.

\section{Conclusions}

It is possible to reconstruct jets up to a large transverse momentum in heavy ion collisions.  A large fraction of the jet energy can be measured as seen by the closeness of nuclear modification factors to 1.  However new physics effects such as momentum dependence of relative quark and gluon sub-processes to inclusive  jet production in the presence of  quark and gluon plasma  and the initial state effects should be considered when interpreting and comparing these results with model calculations \cite{salurQM}.    Some other contributions like the EMC effect might be playing a major role in the relative suppression or enhancement of nuclear modification factors at large momentum \cite{emc}. Implication of jet broadening is observed when comparing different jet definitions with various resolution parameters and  recoil jets of the di-jet coincidence measurements in p+p and Au+Au systems.   In order to study the effects of jet quenching quantitatively, model calculations are required.  Monte-Carlo based simulations of partonic level jet quenching in medium such as Jewel \cite{jewel}, Q-Pythia \cite{qpythia} and YaJEM \cite{yajem} and complementary analytic  calculations \cite{vitev,borghini} recently became available. New robust QCD jet observables that are unaffected by the treatment of hadronization resulting into additional uncertainities need to be explored experimentally to confront these calculations.


\end{document}